\documentstyle[aps,epsf,preprint,pre,eqsecnum]{revtex}
\def\myfigure#1#2{{\leftskip=0.000753\textwidth
\rightskip\leftskip\small
\begin{figure}\baselineskip=14pt plus 2pt minus 1pt
\centerline{#1}\nobreak\smallskip\nobreak #2\end{figure}}}
\global\firstfigfalse
\draft
\newcommand{\bea}{\begin{eqnarray}}     
\newcommand{\eea}{\end{eqnarray}}       
\newcommand{\be}{\begin{equation}}      
\newcommand{\ee}{\end{equation}}
\newcommand{\nn}{\nonumber}     
\newcommand{\rperp}{{\bf r}_{\perp}}
\newcommand{\kperp}{{\bf k}_{\perp}}
\newcommand{\zhat}{{\hat{\bf z}}}
\newcommand{\kperphat}{\hat{\bf k_{\perp}}}
\newcommand{\h}{h(\rperp,t)}
\newcommand{\psid}{\psi(\rperp,t)}

\newcommand{\utilda}{\tilde u}
\newcommand{\rhoplus}{{\rho^{+}(\utilda,t)}} 
\newcommand{\rhominus}{{\rho^{-}(\utilda,t)}} 
 
\newcommand{\wup}{w^{\uparrow}}
\newcommand{\wdn}{w^{\downarrow}}
\newcommand{\nup}{n^{\uparrow}}
\newcommand{\ndn}{n^{\downarrow}}
\newcommand{\pt}{\partial_t}
\newcommand{\vz}{v_{z}(\rperp,0,t)} 
\newcommand{\vel}{{\bf v}({\bf{r}},t)}
\newcommand{\pr}{P({\bf{r}},t)}
\newcommand{\dFdh}{\frac {\delta F}{\delta h}}
\newcommand{\dFdpsi}{\frac {\delta F}{\delta \psi}}

\newcommand{\fh}{{\bf f_h}({\bf{r}},t)} 
\newcommand{\rp}{{\bf {r}}^{\prime}}
\newcommand{\tp}{t^{\prime}}
\newcommand{\F}{{\bf{F}}({\bf{r}},t)}
\newcommand{\hb}{\overline{H}}
\newcommand{\eqker}{4 \eta k_{\perp}} 
\newcommand{\cG}{{\cal{G}}}

\newcommand{\cGaw}{{\cal{G}}^{aw}}
\newcommand{\kpd}{k_{\perp}d}
\newcommand{\tilfz}{\tilde{f_z}} 
\newcommand{\fml}{F_{l}^{m}} 
\newcommand{\fwz}{F_{z}^{w}} 
\newcommand{\fwl}{F_{l}^{w}} 
\newcommand{\R}{{\bf R}} 

\begin{document}

\title{Two-Component Fluid Membranes Near Repulsive Walls:\\
Linearized Hydrodynamics of Equilibrium\\
and Non-equilibrium States}

\author{Sumithra Sankararaman\footnote{Email:sumithra@imsc.ernet.in}
and Gautam I. Menon\footnote{Email:menon@imsc.ernet.in}}
\address{The Institute of Mathematical Sciences,\\
C.I.T. Campus, Taramani, Chennai 600 113,\\
India.}
\author{P.B. Sunil Kumar\footnote{Email:sunil@physics.iitm.ac.in}}
\address{Department of Physics,\\
Indian Institute of Technology Madras, \\
Chennai 600 036,\\
India.}

\date{\today}
\pagebreak
\maketitle
\begin{abstract}

We study the linearized hydrodynamics of a
two-component fluid membrane near a repulsive
wall, via a model which incorporates curvature-
concentration coupling as well as hydrodynamic
interactions. This model is a simplified version
of a recently proposed one [J.-B. Manneville {\it
et al.} Phys. Rev. E {\bf 64}, 021908 (2001)] for
non-equilibrium force-centres embedded in fluid
membranes, such as light-activated
bacteriorhodopsin pumps incorporated in
phospholipid egg phosphatidyl choline (EPC)
bilayers. The pump/membrane system is modeled as
an impermeable, two-component bilayer fluid
membrane in the presence of an ambient solvent,
in which one component, representing active
pumps, is described in terms of force dipoles
displaced with respect to the bilayer midpoint.
We first discuss the case in which such pumps are
rendered {\em inactive}, computing the mode
structure in the bulk as well as the modification
of hydrodynamic properties by the presence of a
nearby wall. These results should apply, more
generally, to equilibrium fluid membranes
comprised of two components, in which the effects
of curvature-concentration coupling are
significant, above the threshold for phase
separation.  We then discuss the fluctuations and
mode structure in steady state of active
two-component membranes near a repulsive wall. We
find that proximity to the wall {\em smoothens}
membrane height fluctuations in the stable
regime, resulting in a logarithmic scaling of the
roughness even for initially tensionless
membranes. This explicitly non-equilibrium result
is a consequence of the incorporation of
curvature-concentration coupling in our
hydrodynamic treatment. This result also
indicates that earlier scaling arguments which
obtained an increase in the roughness of active
membranes near repulsive walls upon neglecting
the role played by such couplings, may need to be
reevaluated.

\end{abstract}
\vskip1cm
\pacs{PACS : 87.16.Dg, 05.20.Jj, 05.40.-a}

\section{Introduction}\label{sec:intro}

Amphiphilic molecules, in polar solvents such as
water, form symmetric bilayer membrane phases at
sufficient concentration
\cite{nelsonpiranweinberg,lipowskynature,seifertreview,lipo95d}.
Such membranes are typically {\em fluid} in nature,
with short-ranged positional order, although more complex 
forms of ordering are
possible\cite{complex1,complex2,complex3}.  The
equilibrium conformations of such single-component
membranes are governed by the energy cost for bending
the bilayer, provided the surface tension can be
neglected, as is the case if the membrane is
self-assembled \cite{lipo95d,gg:gomp97c,lipo95b}.  Our
understanding of the static and dynamic properties of
such single-component membrane systems in equilibrium
is now fairly detailed
\cite{seifertreview,helfrich,goulian93,park,sriramprostlubensky,pfeuty,fdavid,sleibler,seifertlanger}.

More complex multi-component membranes have also been synthesized and
studied \cite{lipo95b,ring88,sunil99,sunil01}.  Such membranes are
especially important in the biological context, since cell membranes
are often usefully idealized as a bilayer {\em fluid-mosaic} comprised
of over a hundred different types of lipid and protein
constituents\cite{singernicolson,brucealberts}.  Biological membranes,
however, often have additional complicating features: An underlying
network of cross-linked proteins anchored to the bilayer and
associated with the cell cytoskeleton typically lends such membranes a
non-vanishing (although small) shear modulus \cite{brucealberts}.
Another aspect, specific to the biological context, is the presence of
non-equilibrium driving forces: An important class of trans-membrane
proteins are the ``ion pumps'', molecules which consume energy derived
from ATP hydrolysis, electro-chemical gradients or light, and undergo
conformational changes.  Such pumps maintain electro-osmotic potential
gradients across the cell membrane by controlling the flow of ions
such as $K^+$ and $Na^+$.  Since energy must be supplied externally,
the driving of the pump is a process which occurs out of thermal
equilibrium.  Thus, the situation of active pumps diffusing in a fluid
membrane matrix is an intrinsically non-equilibrium problem whose
behavior represents a novel class of non-equilibrium steady states.

The appropriate statistical description of active or ``driven''
biological membrane systems has attracted recent experimental and
theoretical attention\cite{srirammadan}. It has been realized that
non-equilibrium behavior may underlie aspects of bio-membrane dynamics
\cite{levinkorenstein,tuvia} previously attributed purely to
equilibrium thermal fluctuations, such as the ``flicker'' phenomenon
in erythrocytes\cite{brochardlennon}.  Recent micropipette experiments
on light-activated bacteriorhodopsin (BR) pumps incorporated in
phospholipid (EPC) bilayers find that area fluctuations in such active
membrane systems can be phenomenologically described in terms of an effective
temperature which exceeds the true physical temperature by up to a
factor of 2 \cite{mannbasslevyprost,sriramprostmannbass}. In these
systems, the transition between active and inactive states can be
easily (and reversibly) tuned, providing a remarkable window into the
dynamics of steady states away from equilibrium
\cite{sriramprostmannbass,prostmannbruinsma,prostbruinsma,sriramprosttoner}.

Analytical calculations which deal with the complex nature of a
typical biological membrane {\it in toto} are difficult, if not
impossible. We may, however, hope to gain useful intuition by working
with simpler models.  We idealize a typical biological membrane here
as composed of principally {\em two} different types of molecular
constituents, the lipids which constitute the bulk of the bilayer
membrane and proteins which diffuse freely on the membrane
surface. Such simple lipid/pump protein systems can be reconstituted
and studied {\it in vitro}, as in the micropipette experiments
referred to above. Our results should apply, most directly, to
such experiments.

The question we address in this paper is the
following: How are the fluctuations of such a
two-component membrane\cite{clarify}, both in
equilibrium and out of equilibrium, affected by the
presence of a nearby repulsive wall?  Fluctuations of
a single-component membrane bounded on one side by a
repulsive wall have been considered in earlier work
\cite{seifertreview,seifertpre}.  More recently, a
similar problem for active membranes has been studied
in an influential paper by Prost, Manneville and
Bruinsma \cite{prostmannbruinsma,prostbruinsma}, who
suggest that steady-state fluctuations of
non-equilibrium membranes near repulsive walls can be
{\em amplified} substantially relative to the
equilibrium case. This amplification is studied via a
scaling treatment which incorporates non-linear
effects. Our starting model differs from theirs in
several respects, in particular in the way we describe
activity, in our incorporation of
curvature-concentration coupling, as well as in our
evaluation of the relative importance of permeative
and hydrodynamic effects at the length scales accessed
by typical experimental probes.  We study the {\em
linearized} hydrodynamics of such active pump-membrane
systems in this paper, ignoring the role of non-linear
effects.  We will describe the similarities as well as
the differences in the results we obtain.

We answer the question posed in the previous paragraph first through a
calculation of the correlations and linearized mode structure in
thermal equilibrium of a model impermeable two-component fluid
membrane, incorporating hydrodynamic interactions.  This amounts, in
terms of the model for biological membranes discussed above, to
assuming that the pump proteins are ``passive'' constituents of the
membrane.  While motivated here in a biological context, such models
are also useful outside this specific context.  Physics similar to
that described in the case where the pumps are rendered passive should
be generic to two-component systems composed of (i) a ``symmetric''
(lipid) constituent forming an up-down symmetric bilayer phase in
isolation, and (ii) an asymmetric (protein) constituent, in which
either the shape of the molecule or its location with respect to the
bilayer favors a local spontaneous curvature for the bilayer.  For
concreteness, we will often use the terms lipid and protein when
referring to constituents of types (i) and (ii), following our earlier
discussion, but stress that the results obtained here in the {\em
equilibrium} case are also applicable more generally.

For protein molecules which lack up-down symmetry, we
may distinguish between molecules oriented parallel to
the local normal and those which are oriented
antiparallel to the normal.  We label these proteins
as ``+'' and ``-'' for convenience; the relatively
small flip-flop rate for protein transfer across
leaves of the bilayer ensures that these labels are
conserved across short and intermediate time scales.
The difference between the
local densities of + and - proteins defines a ``signed '' protein
density field $\psi$.
To describe the incorporation of such asymmetric
proteins into the lipid bilayer, we model the protein
as a rigid rod of length $\wup+\wdn$. The protein is
taken to situate itself asymmetrically with respect to
the bilayer midpoint; a section $\wup$ of the protein
lies on one side of the bilayer midpoint while a
section $\wdn$ lies on the other side. This relatively
simple model of the asymmetry is convenient for
analytic computation.

To lowest order, the two-component character of the membrane manifests
itself via the existence of a curvature-concentration coupling --
fluctuations in $\psi$
influence the mean curvature in that region (see Fig.  1).  We will
work with balanced membranes for which $\langle\psi \rangle = 0$.
For inactive pumps, corresponding to the equilibrium case,
the curvature-concentration coupling term in the free energy effectively
shifts the bending rigidity of the membrane from $\kappa$ to
$\kappa_e = \kappa -{(\kappa \hb)}^2/{\chi_0}^{-1}$. Here $\hb$ is the
coefficient coupling curvature to concentration and ${\chi_0}^{-1}$ an
inverse compressibility \cite{sriramprosttoner}.

In closed vesicular structures found in biological contexts, while the
bilayer constituting the vesicle may be symmetric, chemical and
physical environments within and outside the vesicle can differ
considerably.  Thus, a pump protein can, in principle, distinguish the
side of the bilayer exposed to the inner volume from the side which is
exposed to the outer volume.  For a large enough vesicle, flat on the
large length scales of relevance to a hydrodynamic calculation, the
pump molecule can thus tell {\em up} from {\em down}, where {\em up}
and {\em down} are defined arbitrarily with respect to the average
position and orientation of the bilayer midpoint.  As an extreme
limit, we may consider the case in which the pump molecules
preferentially insert always on the same (say {\em up}) side, so that
a section $\wup$ of the protein molecule always lies {\em above} the
bilayer midpoint while a section $\wdn$ lies below.

A recent study of active membranes works with this model of the
architecture of a membrane-pump system, modeling activity in terms of
dipole forces associated with the ends of the pump, as we detail below
\cite{sriramprostmannbass}. The associated forces may point either
outward at the ends ({\em up} pumps) or inward at the ends ({\em down}
pumps); see Fig. 2. In this model, up and
down pumps are physically distinguishable. The main
results presented here are appropriate to this model.

Note that this distinction between ``above'' and ``below'' is
artificial for strictly symmetric bilayer membranes, in which pump
proteins may insert on either leaf of the bilayer with equal
probability.  In an Appendix, we discuss results for an alternative
model, in which the protein attachment respects the ``up-down''
symmetry of the bilayer; the center of mass of the protein can be
displaced either above or below the bilayer mid-plane as shown in
Fig. 3.  The head and tail distances from the bilayer
midpoint are fixed at $w_l$ and $w_s$ respectively. In this model, up
pumps and down pumps are physically indistinguishable; their
nomenclature follows from the side of the bilayer on which they are
inserted.

We will present results for {\em non-equilibrium}
membranes in the bulk, in which case the pumps are
active, and for the case in which fluctuations of such
an active membrane are bounded on one side by a
repulsive wall. This study complements results by
Prost, Manneville and Bruinsma for a model in which
curvature-concentration coupling is ignored, an effect
explicitly incorporated in later work
\cite{sriramprosttoner}. In the active case, the pumps
exert forces on the membrane and the surrounding
fluid. Since these forces are internal forces, they
must cancel when integrated over the size of the
pump\cite{thanks}.  Following Manneville {\it et al.}
in Ref.\cite{sriramprostmannbass}, we consider the
simplest model for pumps consistent with this
requirement: the pumps are taken to be dipole force
centres with positive and negative force centres
located asymmetrically with respect to the midpoint of
the bilayer.

Our other major assumptions are the following: We
assume that our observations are conducted on scales
such that the membrane may be considered to be {\em
impermeable} to the solvent in which it is embedded.
This assumption follows from the work of Manneville
{\it et al.} who observe that for active terms 
permeative effects can
be neglected {\it vis. a vis\/} hydrodynamic
ones over a substantial range of length scales upto
microscopic ones. We derive our results under the
assumption that the active membrane is impermeable 
and extend this assumption to the case when the membrane
is in the passive state, to smoothly interpolate 
between results in the two regimes.
(In the Rouse or free draining limit,
where the membrane dynamics decouples from that of the
fluid, the membrane is implicitly permeable). We take
the solvent to be incompressible and work at small
Reynolds number, neglecting the inertia of the fluid.
We use linearized equations of motion for the
hydrodynamic velocity field, corresponding to the
Stokes limit of the Navier-Stokes equations.

We now summarize our principal results.
For {\em equilibrium},
impermeable, two-component membranes bounded on one side by a 
wall placed at a distance $d$ from the membrane plane, we 
obtain the following modified mode structure to leading order,
\be 
\omega_1 \approx \cG (\sigma k_{\perp}^2 + \kappa_{e} k_{\perp}^4)\,\,\,\, , \,\,\,\, 
\omega_2 \approx D k_{\perp}^2.
\nn 
\ee
In our notation,  $D$ is the diffusion coefficient
of the proteins on the membrane, $\sigma$ the surface tension of the
membrane, $\kappa$ the bending rigidity of the membrane and $\eta$
the viscosity of the fluid.  Here,
\bea
\cG (k_{\perp},k_{\perp}d) &\sim& \left\{ \begin{array}{ll}
                d^3 k_{\perp}^2 / 12 \eta & \mbox{if $k_{\perp} \ll 1/d$},
\\ 
                1/\eqker & \mbox{if $k_{\perp} \gg 1/d$}. 
           \end{array} \right.
\eea
is a``crossover function'', which
describes how
membrane fluctuations interpolate between a regime
where physical interactions between the membrane and
the wall cannot be neglected, and a regime where such
fluctuations do not feel the presence of the wall\cite{seifertpre}.
In the $k_{\perp} \gg 1/d$ limit, the above result
reduces to one derivable in the bulk case,
as is physically sensible. 

To study {\em non-equilibrium} membranes
we model active pumps, following Manneville
{\it et al.},  as ``asymmetric" force dipoles with force-centres
located at distances $\wup$ above the bilayer midpoint and $\wdn$
below the bilayer midpoint. These distances are, in general,
unequal. The magnitude of the force exerted at each of these
force-centres is $f$.  The pumps diffuse on the membrane and are
characterized by a mobility $\Lambda$.

In the presence of a wall, for non-equilibrium membranes, 
it has been argued that static height fluctuations are {\it amplified} 
and an interesting speculation is
that such an amplification may, in fact, speed up the process of
membrane fusion when two {\em non-equilibrium} membranes are brought
close together \cite{prostmannbruinsma}.  When the wall is
at a distance $d$ from the membrane, our calculation yields (for a
tensionless membrane)
\bea 
\omega_1 &\approx& \frac{f \cGaw \kappa \hb}{{\chi_0}^{-1}} k_{\perp}^2 
+ \cG \kappa_{e} k_{\perp}^4, \nn \\ \omega_2 &\approx& D
k_{\perp}^2 + \frac{\kappa \hb (\kappa \hb \cG k_{\perp}^2 - f
\cGaw)}{{\chi_0}^{-1}} k_{\perp}^2.  
\eea
where $\cGaw$ is the crossover function 
\bea 
\cGaw (k_{\perp},d,\wup,\wdn) &\sim& \left\{ \begin{array}{ll}
         d {\wdn}^2 k_{\perp}^2 /4 \eta & \mbox{if $k_{\perp} \ll 1/d$},
\\ 
                \Omega(k_{\perp},\wup,\wdn)/\eqker & \mbox{if $k_{\perp}
\gg 1/d$}. 
           \end{array} \right.
\eea
$\Omega(k_{\perp},\wup,\wdn)$ is proportional to the degree of
asymmetry $(\wdn-\wup)$. In a regime where fluctuations feel the wall,
non-equilibrium effects do not depend on the asymmetry factor but only
on $\wdn$, as seen above.  Here again, curvature-concentration
coupling plays a crucial role in the manifestation of non-equilibrium
effects in the linearized mode structure. 

We find that in the presence of the wall, the first term
in $\omega_1$ is the most important since it is of order
$O(k_{\perp}^4)$. The existence of this mode depends crucially on the
presence of both activity and curvature-concentration
coupling.  It will be shown later in this paper that
this mode causes active membranes near walls to relax
faster than their equilibrium counterparts and causes
a smoothening of membrane fluctuations;
in the presence of a wall, the roughness of
{\em tensionless}, active membranes averaged over a patch of
size $L \times L$, ($\langle h^2(L)\rangle$), scales
as ln(L) instead of $L^2$.  Thus the wall acts to {\em
smoothen} active membrane fluctuations in a
tensionless membrane by inducing an apparent surface
tension. This result is a purely non-equilibrium one,
since the apparent surface tension vanishes when the
strength of active forces is set to zero.  It also
relies crucially on the existence of a
curvature-concentration coupling, since the apparent
surface tension also vanishes when the curvature and
concentration fields are decoupled. We show that this
effect can be understood in terms of the dynamics of
the fast relaxing $\psi$ field which drives the height
field $\h$ {\it via} activity and
curvature-concentration coupling. 

The outline of this paper is as follows: Section \ref{statics}
describes our basic coarse-grained free energy and reviews the
calculation of the static properties of a two-component membrane with
curvature-concentration coupling. In Section \ref{eqbulk} (a), the
free-draining approximation (Rouse model) is used to discuss the
dynamics.  In Section \ref{eqbulk} (b), we incorporate the effects of
the hydrodynamic velocity field of the solvent. Section \ref{eqwall}
describes the mode structure in the presence of a wall for the
equilibrium problem.  In Section \ref{neqbulk}, we discuss the mode
structure and correlation functions of active two-component
membranes. Section \ref{neqwall} discusses how these properties are
modified by the presence of a wall.  In Section \ref{conclu}, we
summarize the conclusions of this paper. An Appendix discusses the
details of an alternative model for active two-component membranes,
with up-down symmetry.

\section{Statics}\label{statics}

We model the membrane as an
infinitesimally thin two-dimensional surface embedded
in three-dimensional space. Points on this surface are
specified by a three-dimensional vector $\R(\utilda)$,
with components $R_i(\utilda),~i = 1 \ldots 3$.  We
use Monge gauge, valid for nearly flat surfaces, in
which the surface is specified by its
height above a two-dimensional flat plane. This plane
is parameterized via Euclidean
coordinates {\it i.e.} $\utilda = (x,y) \equiv
{\rperp}$; ${\R}(\rperp) = (x,y,h(x,y))$.  In this
gauge, the curvature is $H = -\nabla^2 h$, to lowest
order.

The number densities of the lipid molecules, +
proteins and - proteins are specified through fields
$\rho_\ell$, $\rhoplus$ and $\rhominus$ respectively. These
are the physical densities of the lipids and proteins
on the membrane. The projections of these densities on
the $x-y$ plane are obtained by multiplying the
physical density fields by the metric factor
$\sqrt{1+{(\nabla h)}^2}$. To lowest order in
gradients of $h$, physical densities and projected
densities coincide.  The scalar fields
$\phi(\utilda,t)$ and $\psi(\utilda,t)$ are defined by
$\phi(\utilda,t) = \rhoplus + \rhominus$ and
$\psi(\utilda,t) = \rhoplus - \rhominus$.

The free energy for the membrane-protein system 
 consists of a part arising from protein and lipid
densities (the matter part) and another which arises as a consequence
of membrane fluctuations about the flat state.  Given
equilibrium concentrations of lipid and protein ($\rho_0$),
$\langle\phi(\utilda,t)\rangle = \langle\rhoplus + \rhominus\rangle =
\phi_0$ , $\langle\psi(\utilda,t)\rangle = \langle\rhoplus -
\rhominus\rangle = \psi_0$ and fixing
$\rho \simeq \rho_0$ and $\phi \simeq \phi_0$, 
the appropriate gauge-invariant free energy of the membrane with
asymmetric proteins is
\bea
F &=& \int d^2 \utilda \sqrt{g} [\sigma + 
\frac{1}{2} \chi_0^{-1} (\psi-\psi_0)^2 ] + 
\frac{1}{2} \kappa \int d^2 \utilda \sqrt{g} (H-H_0)^2. 
\eea
where $\sigma$ is the surface tension and $\kappa$ is the bending
rigidity. To lowest order we take $ H_0=\hb \psi$.
To leading order in gradients of $h$, we have
\be
\label{free}
F = \frac{1}{2}\int d^2\rperp[\kappa {(\nabla^2 \h)}^2+\sigma
{(\nabla \h)}^2 - 2\hb \kappa \psid \nabla^2 \h \nn \\
+\chi_0^{-1}
{\psid}^2],
\ee
yielding a local force density of
${\bf f}(\rperp) = -\delta F/\delta h \Big|_\psi$. The 
local force density is evaluated at constant density \cite{compress}.

The partition function is\cite{cailubenskynelsonpowers}
\be
Z = \int {\cal {D}}[h] {\cal {D}}[\psi] e^{-\beta F[h,\psi]},
\ee
which yields, on integrating out the $\psi$ field, an effective
free energy for the height field
$F_{eff}[h]$,
\be 
F_{eff}[h] = \frac{1}{2} \sum_{k_\perp} (\sigma k_\perp^2 + \kappa
k_\perp^4) |h(k_\perp)|^2 - \frac{(\kappa \hb)^2}{{\chi_0}^{-1}}
k_\perp^4 |h(k_\perp)|^2.  
\ee
The equipartition theorem yields the equilibrium correlation function
for the height field fluctuations (about the equilibrium flat, ``mixed'' phase) as
\be
\label{statcorr}
\langle h^{*}(\kperp)h(\kperp)\rangle = \frac{k_BT}{(\sigma k_{\perp}^2
+\kappa_{e} k_{\perp}^4)},
\ee
where
$\kappa_{e} = \kappa - {(\kappa \hb)}^2/{\chi_0}^{-1}$.
The asymmetric components act to reduce the rigidity modulus from
$\kappa$ to $\kappa_{e}$ as a consequence of curvature-concentration
coupling \cite{sriramprosttoner}, an effect easily understood on
physical grounds. We assume that $\hb$ is small enough to keep
$\kappa_{e}$ positive, in which case the membrane is linearly stable.
Instabilities arising at equilibrium due to large values of $\hb$ have
been discussed in \cite{unstableequil}.

\section{Mode structure in the bulk}\label{eqbulk}

We now study the dynamics of a two-component membrane with
curvature-concentration coupling and suspended in a solvent.  We will
first present results in the Rouse model limit in which the background
fluid drains freely through the membrane and effects due to the
hydrodynamics of the background fluid velocity field can be ignored
\cite{doiedwards}.

\subsection{Rouse Model} In
this model the fluid velocity field decouples from the membrane height
field and $\psi$.  The membrane is implicitly permeable and the height
field fluctuations are governed by a balance between the elastic
forces and local (permeative) friction.  An equation of the form
\be 
\label{rouseheight}
\pt \h = -\Gamma \dFdh + f_{m},
\ee
follows, where $\Gamma$ is a kinetic coefficient and $f_{m}$ is a
thermal equilibrium noise whose statistics ensures that the static
height-height correlation function is Eq.\ (\ref{statcorr}).

The proteins diffuse freely on the surface of the membrane, Since
$\psi$ is locally conserved, it obeys an equation of continuity of the
form
\be
\label {diffusion}
\pt \psid = \Lambda \nabla^2 \dFdpsi + \bf{\nabla}\cdot \bf{f_{\psi}},
\ee
with $\Lambda = D/{\chi_0}^{-1}$, D is the diffusion coefficient of
the proteins on the membrane. A 
contribution due to the in-plane current of the pumps has been
dropped; this is typically small and can be neglected to linear order.  
This term contributes if we go beyond the strictly linear treatment
by replacing $\psi^2$ by $\langle \psi^2 \rangle$, yielding
a convective term which gives rise to 
waves whose speed is independent of the wavevector
\cite{srirammadan,sriramprosttoner}. 
The last term is a conserving Gaussian noise with
$\langle \bf{f_{\psi}}(\rperp,t)\rangle = 0$
and correlations
$\langle f_{\psi i}(\rperp,t)f_{\psi j}(\rperp^{\prime},\tp)\rangle = 
2\Lambda k_BT\delta_{ij}
\delta(\rperp-\rperp^{\prime})\delta(t-\tp)$.

Equations (\ref{rouseheight}) and (\ref{diffusion}) can
be written as
\bea
\label{rousematrix}
\pt \left( \begin{array}{c}
            h(\kperp,\omega) \\ 
            \psi(\kperp,\omega)
            \end{array} \right) &=&    -\left( \begin{array}{cc}
                \Gamma (\sigma k_{\perp}^2+\kappa
k_{\perp}^4)& \Gamma \kappa \hb k_{\perp}^2\\ 
                \Lambda \kappa \hb k_{\perp}^4 &
\Lambda \chi_0^{-1} k_{\perp}^2 
                            \end{array} \right)
                        \left( \begin{array}{c}
                        h(\kperp,\omega) \\ 
                        \psi(\kperp,\omega)
                        \end{array} \right) 
+ \left( \begin{array}{c}
                f_{m}\\ 
                i \kperp.\bf {f_{\psi}} 
                \end{array} \right). 
\eea
Solving these coupled equations yields
\bea 
\label{rousecor} 
\langle h^{*}(\kperp,\omega)h(\kperp,\omega)\rangle = \frac{ \langle f_{m}
f_{m} \rangle (\omega^2 + {\tau_{\psi}^{-1}}^2) + 2 {(\Gamma \kappa
\hb)}^2 \Lambda k_B T k_{\perp}^6}{{(-\omega^2 + \tau_{h}^{-1}
\tau_{\psi}^{-1} - \Lambda \Gamma {(\kappa \hb)}^2 k_{\perp}^6)}^2 +
\omega^2 {(\tau_{\psi}^{-1}+\tau_{h}^{-1})}^2}, \eea
where
$\tau_{h}^{-1} = \Gamma (\sigma k_{\perp}^2 +\kappa k_{\perp}^4)$ and
$\tau_{\psi}^{-1} =  \Lambda \chi_0^{-1} k_{\perp}^2$.

On integrating over $\omega$, this correlation function should reduce
to the same form as Eq.\ (\ref{statcorr}). This constrains the
equilibrium thermal noise correlations to be the following:
\be
\langle f_{m}(\rperp,t)f_{m}(\rperp^{\prime},\tp)\rangle = 2\Gamma k_BT
\delta(\rperp-\rperp^{\prime})\delta(t-\tp).
\ee
In addition, we impose $\langle f_{m}(\rperp,t)\rangle = 0$.

To obtain the mode structure, we drop the noise terms in Eq.\
(\ref{rousematrix}) and find the eigenvalues of the matrix. This
yields
\bea
\omega_1,\omega_2 &=& \frac{1}{2} \left(\Gamma (\sigma
k_{\perp}^2 + \kappa k_{\perp}^4) + D k_{\perp}^2 \right) \pm
\frac{1}{2}\left(\sqrt{{(\Gamma (\sigma k_{\perp}^2 + \kappa
k_{\perp}^4) + D k_{\perp}^2)}^2 - 4 D \Gamma k_{\perp}^2 (\sigma
k_{\perp}^2 + \kappa _{e} k_{\perp}^4)}\right)\nn 
\eea
In the absence of curvature-concentration coupling these modes
decouple and are $\omega_1, \omega_2$ = $\Gamma (\sigma k_{\perp}^2
+\kappa k_{\perp}^4)$, $D k_{\perp}^2$, the dispersive modes of a
tense membrane \cite{cailubenskypre}.  For tensionless
membranes, the presence of curvature-concentration coupling modifies
the mode structure to
\be 
\omega_1 \approx \Gamma \kappa_{e} k_{\perp}^4, \omega_2 \approx D
k_{\perp}^2 + \frac{{(\kappa \hb)}^2}{{\chi_0}^{-1}} \Gamma
k_{\perp}^4. \nn 
\ee
\subsection{Zimm Model} 
Next, we examine the effect of solvent hydrodynamics on the dynamical
correlation functions of the protein density field and the membrane
height field \cite{doiedwards}.  Since no external forces act on the
(fluid+membrane+proteins) system, momentum conservation requires
\cite{lubenskybook,landauhydrodyn}
\be
\label{momentum}
\pt g_i = -\nabla_j \pi_{ij},
\ee
where ${\bf g}$ is the momentum density and $\pi_{ij}$
is the stress tensor. The stress tensor takes the form
$ \pi_{ij} = P \delta_{ij} + \rho v_i v_j -\eta (\nabla_i v_j + \nabla_j v_i) 
+ {\cal S}_{ij},$
where $\vel$ is the fluid velocity, $\pr$ is the pressure field and
${\cal S}_{ij}$ describes the contribution to the stress tensor
arising from membrane conformations (the $\nabla^2$ in the
above equation is evaluated in 3 dimensions). In the Stokes limit, the
non-linear term can be dropped.  At low Reynolds number, we neglect
the inertial term on the left hand side of Eq.\ (\ref{momentum})
leading to
$\nabla_j \pi_{ij} = 0$, 
and thus the equation governing the hydrodynamic velocity field
\be 
\label{navier} \eta \nabla^2 \vel
-{\bf \nabla} \pr - \F + \fh = 0.  
\ee
$\F$ is the force acting on the fluid due to membrane stresses ${\cal
S}_{ij}$ and $\fh$ is the equilibrium thermal noise present in the
bulk of the fluid. This noise obeys $\langle\fh\rangle = 0$
\cite{sriramprostmannbass,prostmannbruinsma} and is related to the
viscosity via the fluctuation-dissipation theorem.
\be 
\label{hydcorr}
\langle f_{hi}({\bf{r}},t)f_{hj}(\rp,\tp)\rangle = 
2k_BT\eta\{-\delta_{ij}\nabla^2+\partial_i\partial_j\}
\delta({\bf{r}}-\rp) \delta(t-\tp), \nn
\ee   
where $(i,j)$ index the components of the force.  Eqns.\ (\ref
{navier}) are to be solved together with the incompressibility
condition
\be
\label {incompress}
{\bf \nabla} \cdot \vel = 0,
\ee
to obtain the mode structure of an impermeable two-component membrane
in the presence of an incompressible solvent. 

We employ a useful
convention due to Seifert in which a basis spanned by the
$z$-component and the longitudinal and transverse {\em in-plane}
components of the velocity is used \cite{seifertreview}. (This
convention will be useful further on in this paper when we discuss
membrane fluctuations in the presence of a confining wall).  We solve
the above equations to obtain the transverse ($t$), longitudinal ($l$)
and $z$ components of the fluid velocity in terms of the $t$,$l$ and
$z$ components of the force $\bf{F}$ acting on the fluid \cite{components}. 
The $t$ and $l$ components of the velocity lie in the plane of the membrane
and are perpendicular and parallel to $\kperp$, the wavevector describing
fluctuations of the membrane. Some algebra yields these components as \cite{seifertreview}
\bea
\label{planecomp}
v_z(\kperp,z)&=&\frac{1}{4\eta k_{\perp}}\int_{-\infty}^{\infty} dz^{\prime}
e^{-k_{\perp}|z^{\prime}-z|}[(1+k_{\perp}|z-z^{\prime}|)F_z(\kperp,z^{\prime})
+ i k_{\perp}(z^{\prime}-z)F_l(\kperp,z^{\prime})], \nn \\
v_l(\kperp,z)&=&\frac{1}{4\eta k_{\perp}}\int_{-\infty}^{\infty} dz^{\prime}
e^{-k_{\perp}|z^{\prime}-z|}[(1-k_{\perp}|z-z^{\prime}|)F_l(\kperp,z^{\prime})
+i k_{\perp}(z^{\prime}-z)F_z(\kperp,z^{\prime})]
\eea
\bea
\label {tcomp}
v_t(\kperp,z)&=&\frac{1}{4\eta k_{\perp}}\int_{-\infty}^{\infty} dz^{\prime}
e^{-k_{\perp}|z^{\prime}-z|} 2F_t(\kperp,z^{\prime}). 
\eea
Note that the transverse components can be ignored since they do not
couple to the other components. 
We can now calculate $v_z(\kperp,0)$ by supplementing the above
equations with boundary conditions.

In the absence of permeation the membrane is advected by the
fluid. The velocity of any point on the membrane is the $z$-component
of the velocity of the fluid at that point\cite{sriramprostmannbass}.
The equation of motion for the height field of an impermeable membrane
is therefore
\be 
\label {advection} 
\pt \h = \vz, 
\ee 
where $\vz$ is the velocity of the surrounding fluid at the mean
position of the membrane.  The force acting on the fluid is of the
form
\be
{\bf {F}}(\kperp,z) = \delta(z)[-\dFdh \zhat + \fml
\kperphat],
\ee
where $\fml$ is the longitudinal component of the force acting on the
fluid due to the membrane.  The elastic force density due to the
membrane is $-\delta F/\delta h\zhat$ .  We
solve for $v_z(\kperp,0)$ using Eqns.\ (\ref {planecomp}),
together with the boundary condition that the membrane is
incompressible\cite{compress} {\it i.e.} the in-plane divergence 
of the velocity field vanishes at the membrane. 
This implies that $v_l(\kperp,0) = 0$.  
This boundary
condition gives $\fml=0$, yielding
\be
v_z(\kperp,0) = -\frac{1}{\eqker} \dFdh.
\ee
The bulk thermal noise present in the fluid also
contributes to $v_z(\kperp,0)$. This contribution is denoted by
$f_z$, with
\be
f_z(\kperp,0) = \int \frac{dk_z}{2\pi\eta k^2}(\delta_{jz}-\frac{k_jk_z}{k^2}) 
{[f_h({\bf k},t)]}_j.
\ee
We can use the noise correlation function given in Eq.\ (\ref
{hydcorr}) to show that the noise $f_z$ has zero mean and that its variance
in Fourier space is $2k_BT/4\eta k_{\perp}$.

Finally, Eq.\ (\ref {advection}) becomes
\be
\pt h = -\frac{1}{\eqker} \dFdh + f_z(\kperp,0,t).
\ee
The density difference field obeys Eq.\ (\ref {diffusion}).  Eq.\
(\ref {advection}) and Eq.\ (\ref {diffusion}) can then be written as  
\bea 
\label{zimmmatrix}
\pt \left(
\begin{array}{c}
            h(\kperp,\omega) \\  \psi(\kperp,\omega) \end{array} \right) &=&
            -\left( \begin{array}{cc}
                                                \frac{\sigma
                                                k_{\perp}^2+\kappa
k_{\perp}^4}{\eqker}& \frac{\kappa \hb k_{\perp}^2}{\eqker}\\ 
                                                \Lambda \kappa \hb
                                                k_{\perp}^4 &
\Lambda \chi_0^{-1} k_{\perp}^2
                                            \end{array} \right) \left(
                                        \begin{array}{c}
                                                h(\kperp,\omega) \\ 
                                                \psi(\kperp,\omega)
                                                \end{array} \right) 
+ \left( \begin{array}{c}
                f_z(\kperp,\omega)\\ 
                i \kperp.\bf {f_{\psi}} 
                \end{array} \right). 
\eea 
Solving these equations gives
\bea 
\langle h^{*}(\kperp,\omega)h(\kperp,\omega)\rangle = \frac{ (2
k_B T/4 \eta k_{\perp})(\omega^2 + {\tau_{\psi}^{-1}}^2) + (k_B T
\Lambda {(\kappa \hb)}^2 /8 \eta^2) k_{\perp}^4}{{(-\omega^2 +
\tau_{h}^{-1} \tau_{\psi}^{-1} - (\Lambda {(\kappa \hb)}^2/4 \eta)
k_{\perp}^5)}^2 + \omega^2 {(\tau_{\psi}^{-1}+\tau_{h}^{-1})}^2}.
\eea
The parameters appearing in the above equations are 
\be
\tau_{h}^{-1}=\frac{1}{\eqker}
(\sigma k_{\perp}^2 +\kappa k_{\perp}^4), 
\tau_{\psi}^{-1} = \Lambda \chi_0^{-1} k_{\perp}^2.\nn
\ee
Integrating this result over $\omega$ recovers for us the equal time
correlation function for height fluctuations about the equilibrium,
flat, ``mixed'' phase
\be
\label {equicorr}
\langle h^{*}(\kperp,t)h(\kperp,t)\rangle = \frac{k_BT}{(\sigma k_{\perp}^2
+\kappa_{e} k_{\perp}^4)},
\ee
as expected. The roughness of the height field at macroscopic length
scales $L$, defined through $\langle h^2(L)\rangle = \int_{1/L}^{1/a}
d^2k_{\perp} \langle h(k_{\perp}) h^{*}(k_{\perp}) \rangle$ scales as
\bea
\label{equilcorr}
 \langle h^2(L)\rangle = \left\{ \begin{array}{ll}
                        \frac{k_BT}{\kappa_{e}} L^2 & \mbox{if $\sigma=0$},\\
                        \frac{k_BT}{\sigma}ln(L/a) & \mbox{if $\sigma \ne
0$},
                       \end{array}
              \right.  
\eea
where a is a microscopic cut-off length.

The mode structure is then
\bea
\omega_1,\omega_2 &=&
\frac{1}{8 \eta} \left(\sigma k_{\perp} + \kappa k_{\perp}^3 + 4 \eta D
k_{\perp}^2 \right) \pm \frac{1}{8 \eta} \left(\sqrt{{(\sigma k_{\perp}
+ \kappa k_{\perp}^3 + 4 \eta D k_{\perp}^2)}^2 - 16
\eta D (\kappa_{e} k_{\perp}^5 + \sigma k_{\perp}^3)} \right).\nn
\eea
In the absence of curvature-concentration coupling, $\omega_1,\omega_2
= (\sigma k_{\perp}+\kappa k_{\perp}^3) /4 \eta,D k_{\perp}^2$, {\it
i.e.} the decoupled modes of the membrane height field and the protein
concentration field in the presence of hydrodynamic correlations.
For tensionless membranes with non-zero
curvature-concentration coupling, the modes are
\be
\omega_1 \approx \frac{\kappa_{e}}{4 \eta} k_{\perp}^3, 
\omega_2 \approx D k_{\perp}^2 +\frac{{(\kappa \hb)}^2}
{4 \eta {\chi_0}^{-1}} k_{\perp}^3. \nn 
\ee

\section{Mode Structure in the presence of a wall}\label{eqwall}

Consider a fluid bilayer membrane with shape
asymmetric constituents, suspended in a fluid bounded
on one side by a wall. Our model for the membrane-pump
system follows that of Manneville {\it et al.}. In our
co-ordinates, the membrane fluctuates about the $z=d$
plane and the wall is at $z=0$. The membrane is kept
at distance $d$ from the wall by a repulsive
interaction with the wall and a suitable external
pressure which prevents it from floating away into 
the bulk. 

The relaxation rate of an equilibrium
membrane is reduced near a wall; in tandem, the equilibrium noise
fluctuations are required to have a reduced amplitude as a consequence
of the fluctuation-dissipation theorem \cite{prostmannbruinsma}. On
short-length scales, however, membrane fluctuations do not see the
wall and the height-height correlation function is identical to that
in the bulk.  This can be used to determine the form of thermal
correlation functions in the presence of the wall.

We work in the linear regime where $\langle h^2(L_c)\rangle \approx
d^2$, defining a collision length $L_c$.  Beyond this length the
membrane senses the presence of the wall and the non-linearities
neglected in our treatment become important; $1/L_c$ is the smallest
wave vector above which the linearization condition $|h(\rperp,t)| \ll d$
is valid.  From Eq.\ (\ref{equilcorr}) we can estimate the collision
length of an equilibrium membrane near a wall to be
\bea
 L_c(d) = \left\{ \begin{array}{ll}
                        \sqrt{\frac{\kappa_{e}}{k_BT}} d & \mbox{if
$\sigma=0$},\\ 
                        a~exp(\sigma d^2/k_BT) & \mbox{if $\sigma \ne
0$}.
                       \end{array}
              \right.
\eea

We now solve Eq.\ (\ref{navier}) to obtain the velocity
of the fluid at the position of the membrane.  The
membrane and wall are introduced in Eq.\ (\ref
{navier}) as external forces acting on the fluid, imposing
the required boundary conditions.
The incompressibility of the fluid is used to solve for
the components of the velocity in terms of the
components of the external force.

We express the
components of the velocity in the ``mixed'' basis of
Seifert \cite{seifertreview} as before. 
The force exerted by the
membrane and wall on the fluid is of the form \cite{seifertreview}
\be
{\bf{F}}(\kperp,z)=\delta(z-d)[-\dFdh \zhat+\fml \kperphat]
+\delta(z)[\fwz \zhat + \fwl \kperphat],
\ee
where $\fwz$ and $\fwl$ are the $z$ and longitudinal components of the
force exerted by the wall on the fluid at $z=0$. The quantities
$\fml$, $\fwz$ and $\fwl$ can be evaluated using the boundary
conditions: 
all components of the velocity should vanish at the wall (no-slip
boundary condition) \cite{landauhydrodyn}.  This imposes
\be
v_z(\kperp,0,t) = v_l(\kperp,0,t) =  v_t(\kperp,0,t) = 0.
\ee
These conditions, together with membrane incompressibility
($v_l(\kperp,d,t) =0$), provide the boundary conditions for
the velocity field of the fluid at 
the membrane. This gives $v_z(k_{\perp},d,t) = - \cG \dFdh$. 
The equation of motion for the
height field is then obtained from 
Eqn.\ (\ref{advection})\cite{seifertpre},
\be
\label{conservation}
\pt h(\kperp,t) = -{\cG}\dFdh + \tilfz(\kperp,d,t),
\ee
where 
\bea
\cG (k_{\perp},k_{\perp}d) 
&=& \frac{1}{\eqker} \left(\frac{2(\sinh^2(\kpd)-(\kpd)^2)}{\sinh^2(\kpd)-(\kpd)^2
+\kpd+\sinh(\kpd)\cosh(\kpd)}\right), \nn \\
&\simeq& \left\{ \begin{array}{ll}
                d^3 k_{\perp}^2 / 12 \eta & \mbox{if $k_{\perp} \ll 1/d$}
~~~^{\dag},  \\
                1/\eqker& \mbox{if $k_{\perp} \gg 1/d$}.
           \end{array}
           \right.
\eea
We note that the term marked $\dag$ is incorrect for arbitrarily small
wave vectors. The relevant cut-off is the inverse of the collision
length $L_c$ since linearized calculations are not valid at
wave-vectors smaller than this value.

The contribution to the velocity field at the membrane due to the
noise in the fluid in the presence of the wall is $\tilfz$.  Following
the procedure used in the derivation of Eq.\ (\ref{zimmmatrix}), we
get
\bea 
\label{eqwallmatrix} \pt \left( \begin{array}{c}
            h(\kperp,\omega) \\  \psi(\kperp,\omega) \end{array} \right) =
-\left( \begin{array}{cc}
        \cG(\sigma k_{\perp}^2+\kappa
k_{\perp}^4)& \cG(\kappa \hb k_{\perp}^2) \\
                                        \Lambda \kappa \hb k_{\perp}^4 &
\Lambda \chi_0^{-1} k_{\perp}^2 
                                            \end{array} \right) 
                                            \left(
                                        \begin{array}{c}
                                                h(\kperp,\omega)  \\
                                                \psi(\kperp,\omega)
                                                \end{array} \right) 
+ \left( \begin{array}{c}
                \tilfz(\kperp,\omega)\\ 
                i \kperp.\bf {f_{\psi}} 
                \end{array} \right). 
\eea
This matrix equation is solved to yield
\bea 
\langle h^{*}(\kperp,\omega)h(\kperp,\omega)\rangle = \frac{
\langle \tilfz \tilfz \rangle (\omega^2 + {\tau_{\psi}^{-1}}^2) + 2
k_B T \Lambda {(\cG \kappa \hb)}^2 k_{\perp}^6}{{(-\omega^2 +
{\tau_h^{-1}}^{w} \tau_{\psi}^{-1} - \cG \Lambda {(\kappa \hb)}^2
k_{\perp}^6)}^2 + \omega^2 {(\tau_{\psi}^{-1}+{\tau_h^{-1}}^{w})}^2},
\eea
where ${\tau_h^{-1}}^{w}=\cG(\sigma k_{\perp}^2 + \kappa k_{\perp}^4)$.

The hydrodynamics of the solvent cannot change equilibrium correlation
functions.  Thus, demanding that equal time correlations of the $h$
field are the same as in Eq.\ (\ref {equicorr}), we obtain the noise
correlation function of $\tilfz$ as
\bea
\label{wallnoisecorr}
\langle \tilfz(\kperp,t)\rangle &=& 0, \\
\langle \tilfz(\kperp,t)\tilfz(\kperp^{\prime},\tp)\rangle &=& 2k_BT
\cG(k_{\perp},k_{\perp}d)
\delta(\kperp+\kperp^{\prime}) \delta(t-\tp).
\eea
Note that this reduces to the noise correlation function for the
equilibrium bulk noise when $k_{\perp}>>1/d$, {\it i.e.} when the
membrane fluctuations are not significantly affected by the wall (see
Section III).

We note here that the  equation of motion of the height field 
can be derived in the limit $1/L_c<<k_{\perp}
<<1/d$ by directly imposing impermeability as in \cite{prostmannbruinsma}. 
Since the membrane
is impermeable, the height field is conserved and obeys a continuity equation
$\pt h = - \nabla_{\perp} . j_h$. The 
height ``current'' can be calculated from the lubrication approximation as $j_h =  \frac{-d^3}{ 12 \eta}
\nabla_{\perp} P$, where $P$ is the pressure due to the elastic forces and
the noise.
Using the appropriate expressions for these terms, we recover the equation 
for the height (Eq.\ (\ref{conservation})) in the ``close to the wall'' limit. 

The eigen-modes follow from Eq.\ (\ref{eqwallmatrix}) 
\bea
\omega_1,\omega_2 = \frac{1}{2} \left( D k_{\perp}^2 +\sigma
\cG k_{\perp}^2 \pm \kappa \cG k_{\perp}^4 \right) +
\frac{1}{2}\left(\sqrt{{(\sigma \cG k_{\perp}^2 + D k_{\perp}^2+
\kappa \cG k_{\perp}^4)}^2 - 4 \cG D (\sigma + \kappa_{e} k_{\perp}^2)
k_{\perp}^4} \right)\nn
\eea
In the absence of curvature-concentration coupling the modes
corresponding to the hydrodynamic relaxation of a two-component
membrane near a wall are recovered : $\omega_1,\omega_2 = \cG (\sigma
k_{\perp}^2 + \kappa k_{\perp}^4), D k_{\perp}^2$.  For membranes with
curvature-concentration coupling the modes are modified to
\be
\omega_1 \approx \cG (\sigma k_{\perp}^2 + \kappa_{e} k_{\perp}^4), 
\omega_2 \approx D k_{\perp}^2 + 
\cG \frac{{(\kappa \hb)}^2}{{\chi_0}^{-1}} k_{\perp}^4.
\ee
It is easy to see that these reduce to the expressions derived in the
bulk case, in the limit in which membrane fluctuations are
unaffected by the wall. 

\section{Non-equilibrium Membranes: Mode Structure in Bulk}\label{neqbulk}

We now discuss the dynamic correlations and mode structure of an
impermeable two-component membrane with active force-centres, when
placed far away from confining walls.  For inactive force centres, the
static properties of such a system are given by the free energy of
Eq.\ (\ref{free}).  However, in the presence of non-equilibrium
driving forces, thermodynamic restoring forces must be supplemented by
additional terms.  These are discussed briefly here.

First, we must account for non-equilibrium forces. In the context of
the membrane-pump system, these arise due to the action of the
pumps. However, since pumping refers to forces internal to the system,
the force density exerted on the whole system (membrane $+$ pump $+$
solvent) must integrate to zero on length scales comparable to the
pump size.  A simple representation of this requirement idealizes the
pump as two force-centres of opposite sign but equal magnitude
separated by a distance $\wup+\wdn$ and embedded in the bilayer.  Such
a ``force dipole'' representation was introduced in
Ref.\cite{sriramprostmannbass} and will represent an important part of
our discussion here.

Second, we must account for the non-equilibrium nature of the noise
present in the fluid due to the switching "on" or "off" of the pumps.  
For simplicity, we assume that this noise is delta correlated.
Such an assumption corresponds to
a coarse-graining in time across the typical pump ``dead'' time which
separates two pumping events. The non-equilibrium active noise can now
be absorbed into the correlation function of the equilibrium noise
(Eq.(\ref{hydcorr})). However, note that the ``temperature'' which
appears in this correlator is not the thermodynamic temperature. 
To smoothly interpolate to the equilibrium case, we must
redefine this temperature as an effective temperature which becomes
the true thermodynamic temperature in the absence of activity.

Several of the results in this section overlap with those of
Manneville {\it et al.} (Ref.  \cite{sriramprostmannbass}) in which
the effects of permeation and curvature induced activity on the
dynamics of an active membrane in bulk are considered.  In our model
we assume impermeability from the outset and neglect the effects of
curvature induced activity.  Such a ``minimal'' model makes the
calculation of the correlation functions of a wall-bounded active
membrane more tractable and transparent. 

We begin by writing the equations of motion for the various
fields. The fluid is incompressible and hence obeys
Eq.~(\ref{incompress}).
Since no external forces act on the
(fluid+membrane+proteins) system, momentum is conserved
\cite{lubenskybook,landauhydrodyn} and Eq.(\ref{momentum})
holds.  The momentum current tensor $\pi_{ij}$ is then
\be 
\pi_{ij} = P \delta_{ij} + \rho v_i v_j -\eta (\nabla_i v_j + \nabla_j v_i) + S_{ij}+
A_{ij},
\ee
where $\vel$ is the three dimensional fluid velocity, $\pr$ is the
three dimensional pressure field.  $S_{ij}$ is the contribution to the
stress tensor due to membrane conformations and $A_{ij}$ is the
contribution from the active pumps.  Eqns.\ (\ref {navier}) and (\ref
{incompress}) can be solved as before for the transverse ($t$),
longitudinal ($l$) and $z$ components of the fluid velocity, in terms
of the $t$,$l$ and $z$ components of the force $\F$ acting on the
fluid due to membrane and pump stresses.  The transverse components
will be ignored as before; $\fh$ represents the thermal noise present
in the bulk of the fluid.

We can now calculate $v_z(\kperp,0,t)$ by solving the
equations of motion with appropriate boundary
conditions.  The equation of motion for the height
field in the absence of permeation is $\pt \h = \vz$,
where $\vz$ is the velocity of the surrounding fluid
at the mean position of the membrane.  The membrane
does not feel the forces from the pumps directly
because of the absence of permeation but only
{\it via\/} the forces the pumps exert on the fluid
\cite{sriramprostmannbass,prostmannbruinsma}. The
contribution of the bulk equilibrium thermal noise
present in the fluid to the $z-$component of the fluid
velocity at $z=0$ is $f_z$. The statistics of $f_z$
have been discussed earlier (see Section III).  The
density difference field obeys the diffusion equation
Eq.\ (\ref {diffusion}). As in the equilibrium case,
we work in the linearly stable regime.

The membrane fluctuates about $z=0$.  The force exerted by the
membrane and pumps on the fluid is of the form
\be
{\bf {F}}(\kperp,z^{\prime}) = \delta(z^{\prime})[-\dFdh \zhat+\fml \kperphat]
+f_{pump-fluid} \zhat.
\ee
The force $f_{pump-fluid}$ is 
modelled as a dipolar force density with force-centres located
asymmetrically about the $z=0$ plane.
\be
\label {pumpforceden}
f_{pump-fluid} = f \psid [\delta(z^{\prime}-\wup)-\delta(z^{\prime}+\wdn)],
\ee
where $f$ sets the scale for the active force.

We again solve for the unknown coefficient $\fml$ using 
membrane incompressibility.  This imposes
$v_{l}(k_{\perp},0,t) = 0$ in Eq.\ (\ref {planecomp}).  The
hydrodynamic velocity obtained at $z=0$ using Eq.\ (\ref {planecomp})
enters in the equation of motion for the height field:
\be
\label {actadvect}
\pt h(\kperp,t) = \frac{1}{\eqker}[-\dFdh+f \psi(\kperp,t)
\Omega(k_{\perp},\wup,\wdn)]+f_z(\kperp,0,t),
\ee
where
\bea 
\Omega(\kperp,\wup,\wdn) &=& -e^{-k_{\perp} \wdn}(1+k_{\perp}\wdn) +
e^{-k_{\perp} \wup}(1+k_{\perp}\wup) \\ &=&
\frac{k_{\perp}^2}{2}({\wdn}^2 - {\wup}^2) +
\frac{k_{\perp}^3}{3}({\wdn}^3 - {\wup}^3)+ \ldots. 
\eea
Note that $\Omega(k_{\perp},\wup,\wdn)$ vanishes when $\wup=\wdn$,
implying that an asymmetry in the position of the force-centres
above and below the membrane mid-plane is needed for a
non-zero active (non-equilibrium) force. The
equations of motion are
\bea
\label{noneqbulkmatrix}
\pt \left(
\begin{array}{c}
            h(\kperp,\omega)  \\ \psi(\kperp,\omega) \end{array} \right) =
           -\left( \begin{array}{cc}
                                                \frac{\sigma k_{\perp}^2+\kappa
k_{\perp}^4}{\eqker}& \frac{\kappa \hb k_{\perp}^2-f \Omega(k_{\perp},\wup,\wdn)}{\eqker}\\ 
                                                \Lambda \kappa \hb k_{\perp}^4 &
\Lambda \chi _0^{-1}k_{\perp}^2 
                                            \end{array} \right)
                                            \left(
                                        \begin{array}{c}
                                                h(\kperp,\omega) \\ 
                                                \psi(\kperp,\omega)
                                                \end{array} \right) 
+ \left( \begin{array}{c}
                f_z(\kperp,\omega)\\ 
                i \kperp.\bf {f_{\psi}} 
                \end{array} \right). 
\eea
The correlation function for the height fluctuations of an active
membrane (about the flat, stable, steady state phase) is then
\be 
\label {actcorr} 
\langle h^{*}(\kperp,t)h(\kperp,t)\rangle =
\frac{k_BT}{\eqker}
\frac{(\tau_{\psi}^{-1}+\tau_{h}^{-1}-\tau_{a}^{-1} +\frac{f^2
\Omega^2}{\eqker \chi_0^{-1}})}{(\tau_{\psi}^{-1}
+\tau_{h}^{-1})({\tau_{h}^{-1}}^e+\tau_{a}^{-1})}, 
\ee
where the quantities $\tau_{\psi}^{-1},\tau_{h}^{-1}$ have been
defined earlier. ${\tau_{h}^{-1}}^e$ has the same form as
$\tau_{h}^{-1}$ with $\kappa$ replaced by $\kappa_{e}$. The relaxation
time
\be
\tau_{a}^{-1} =
\frac{1}{\eqker}\frac{\kappa \hb}{{\chi_0}^{-1}} f
\Omega(k_{\perp},\wup,\wdn) k_{\perp}^2. \nn
\ee
We discuss the positivity of the right hand-side of Eq. \ (\ref {actcorr})
after discussing the mode structure of the system.
From the equal time correlation function we can calculate the quantity
$\langle h^2(L)\rangle$. To lowest order we find
\bea
\label{neqhcorrnoten}
\langle h^2(L)\rangle = k_B T L^2 \Big/ \left[\kappa_{e}- 
\frac{\kappa \hb {\cal P} w }{{\chi_0}^{-1}}\right],
\eea
for tensionless membranes and
\bea
\label{neqhcorrten}
\langle h^2(L)\rangle = \frac{k_B T}{\sigma} \ln(L/a),
\eea
for tense membranes. ${\cal P} = \frac{f ({\wup}^2 - {\wdn}^2)} {2 w}
$, where $w$ is the size of the pump, represents the force dipole energy.
It is
estimated to be a few $k_B T$ \cite{sriramprostmannbass}.  These
results imply that on large length scales, pumping activity does not
change the roughness of equilibrium, tensionless membranes but shifts
the {\em temperature} at lowest order to
\bea
\label{efftemp}
T^{eff} = T \left[1+\frac{\kappa \hb {\cal P} w }{ \kappa_{e} 
{\chi_0}^{-1}} \right].
\eea 
For tense membranes both roughness and effective temperature are not
affected by non-equilibrium effects at large length scales since
the divergence of the height correlation function with system size
remains logarithmic.

In the stable regime Eq.\ (\ref{neqhcorrnoten}) is
positive, as discussed later in this section. The correlation
function in  Eq.\ (\ref{neqhcorrnoten}) is calculated from the lowest
order diffusive terms.  
Manneville et al. \cite{sriramprostmannbass} calculate an identical
correlation function but
{\em neglect} these diffusive terms in favor of higher order terms
on the grounds that the diffusion constant is small in the physical
situation. For this reason the effective temperature we calculate in Eq.\
(\ref {efftemp}) 
{\it is not} the effective temperature that Manneville et
al. calculate and compare with experiment.

The height correlation functions we calculate in our linearized
approach (Eqns.\ (\ref{neqhcorrnoten}) and (\ref{neqhcorrten})) show
that the roughness of an active fluid membrane is unchanged in the
bulk. The calculation of  Ref. \cite{prostmannbruinsma} obtains an 
enhanced roughening. The reason is the different nature of the
force densities used in both the calculations.  Prost {\it et al.}
consider a model in which the membrane feels the activity of the pump
{\em only} as a consequence of permeation.  In our calculation, Eq.\
(\ref{actadvect}) written as a Langevin equation is
\be 
\pt h(\kperp,t)
+ \tau_h^{-1} h(\kperp,t)= -\frac{(\kappa \hb
k_{\perp}^2 - f \Omega)}{4 \eta k_{\perp}} \psi + f_z.
\nn 
\ee 
We assume that the pump is always in the active state over the time
scales of our interest.  The forces on the right-hand side of the
above equation arise from three sources: an elastic part (due to the
curvature concentration-coupling term in the free energy), a
asymmetric dipole part (from the activity of the pumps) and the
thermal noise from the fluid.  Since the membrane is impermeable it
does not feel the active forces or thermal noise directly but only
{\it via} the fluid.  We emphasize that in the present model the
membrane senses active forces and thermal noise even when $\lambda_p
=0$, in contrast to the model of 
Prost {\it et. al.}\cite{prostmannbruinsma}.

The modes are 
{\small{
\bea
\omega_1,\omega_2 &=& \frac{1}{8 \eta} \left( \sigma k_{\perp} + 4 \eta D k_{\perp}^2 
+\kappa {k_{\perp}}^{3} \right) \nn \\
&\pm&
\frac{1}{8 \eta} \left(\sqrt{{(\sigma k_{\perp} + 4 \eta D k_{\perp}^2 +\kappa {k_{\perp}}^{3})}^2
-16 \eta (D \kappa_e k_{\perp}^5 
+D \sigma k_{\perp}^3+ f \hb \kappa \Lambda \Omega k_{\perp}^3)} \right).\nn
\eea}}
In the absence of curvature-concentration coupling, non-equilibrium
effects vanish to lowest order and the modes are those for an
equilibrium membrane with hydrodynamic interactions $\omega_1,\omega_2
= (\sigma k_{\perp} + \kappa k_{\perp}^3)/4 \eta, D k_{\perp}^2$.
When $\sigma =0$, curvature-concentration coupling modifies this mode
structure to
\bea
\omega_1 &\approx& \frac{\kappa_{e}}{4 \eta} k_{\perp}^3 + \frac{f \kappa \hb ({\wdn}^2 -{\wup}^2)}
{8 \eta {\chi_0}^{-1}} k_{\perp}^3,\nn \\
\omega_2 &\approx& D k_{\perp}^2 + \frac{ \kappa \hb ( 2 \kappa \hb - f ({\wdn}^2 -{\wup}^2))
k_{\perp}^3}{8 \eta {\chi_0}^{-1}}.
\eea
We see that to lowest order, non-equilibrium effects are apparent
only at non-zero curvature-concentration coupling. We also observe
that the non-equilibrium terms vanish when $\wdn = \wup$.

For stability we require that all the eigenmodes of the system must decay. 
This means that $\omega_1$ and $\omega_2$ are positive. 
To the order above, this imposes
$ \frac{\kappa_e}{\kappa} -1 < \frac {\hb {\cal P} w}{\chi_0^{-1}}
< \frac {\kappa_e}{\kappa}$.
 If $\hb {\cal P}$ is such that this condition is satisfied then the
system is stable independent of the sign of this parameter.
Typical experimental values such as
$\kappa= 10~k_B T$, $\kappa \hb = w~ k_B
T$,$\chi_0^{-1} = w^2 ~k_B T$, $f = 10^{-12} N$, satisfy the above
conditions\cite{comment}.

\section{Non-equilibrium membranes : Mode Structure in the presence of a wall} 
\label{neqwall}

We now study the effect of a wall located at $z=0$ on the
fluctuations of an active membrane located at $z=d$. The forces
exerted on the fluid by the membrane, pumps and the wall are now given
by
\bea
{\bf{F}}(\kperp,z^{\prime})&=&\delta(z^{\prime}-d)[-\dFdh \zhat 
+\fml \kperphat] + f \psid [\delta(z^{\prime}-d-\wup)-\delta(z^{\prime}-d+\wdn)] \zhat
\nn \\
 &+&\delta(z^{\prime})[\fwz \zhat + \fwl \kperphat].
\eea
We need to determine the unknown coefficients $\fml$,$\fwz$ and $\fwl$.
The following three boundary conditions are used, following
Eqns.\ (\ref{planecomp}),
\be
v_z(\kperp,0,t) = v_l(\kperp,0,t) =  v_l(\kperp,d,t) = 0.
\ee
These boundary conditions impose the following
constraints: (i) $v_z(\kperp,z=0) = 0 $, {\it i.e.}
the $z$-component of the velocity vanishes at the wall
(ii) $v_l(\kperp,z=0) = 0$, {\it i.e.} the
$l$-component of the velocity vanishes at the wall,
and (iii) $v_l(\kperp,z=d) = 0$, {\it i.e.}
the membrane located at $z=d$ is incompressible. We
neglect the tangential component as before. 

The values of the three unknown coefficients are
obtained on solving the three simultaneous
equations (Eq. \ (\ref{planecomp})) together with
the three boundary conditions.  These values can
then be used to obtain the velocity of the fluid
at the location of the membrane. The velocity
thus obtained has a part arising from equilibrium elastic
forces and a part arising from non-equilibrium active
forces. These can be separated out, yielding 
the equation of motion for the height
field as
\bea
\label{actwallheight}
\pt h(\kperp,t) &=& v_z(\kperp,d,t) \nn \\
&=&-{\cG}(\dFdh )+\cGaw(f \psi)+ \tilfz(\kperp,t),
\eea
where $\cG$ 
is the usual hydrodynamic kernel obtained for the equilibrium
fluctuations in the presence of a wall and the new {\em
non-equilibrium} term
\bea
\cGaw(k_{\perp},d,\wup,\wdn)  
&\sim& \left\{ \begin{array}{ll}
                 \frac{k_{\perp}^2}{12 \eta} (3 d {\wdn}^2  -2 {\wdn}^3)+O(k_{\perp}^4) +\ldots
 & \mbox{if $k_{\perp} \ll 1/d$},\\ 
\frac{1}{\eqker} \Omega(k_{\perp},\wup,\wdn) & \mbox {if $k_{\perp} \gg
1/d$}. 
           \end{array}
           \right.
\eea
is the hydrodynamic kernel for non-equilibrium fluctuations in the
presence of the wall\cite{prostmannbruinsma}(the expanded form of 
this kernel is given in \cite{noneqkernel}; see also 
Ref.\cite{manndocthesis}). 
In the expressions for $\cG$ and $\cGaw$, the lower cut-off on
$k_{\perp}$ is the inverse of the collision length $L_c$, which we will
calculate towards the end of this section. The membrane-wall distances we
use in our calculations are such that $d>>\wup,\wdn$ and hence
${\wdn}^3$ can be neglected in comparison to $d {\wdn}^2$.  
In the long wavelength limit, $\cGaw$ depends on $\wdn$ and not on
$(\wdn-\wup)$ as in the bulk. When the fluctuations are not affected
by the wall, $\cGaw$ vanishes when $\wup$ equals $\wdn$
since in bulk the directions ``up'' and ``down'' are equivalent. The
non-equilibrium contribution to the force density therefore vanishes.
Thus, in this large membrane-wall distance limit, we recover the
previous situation of active membrane fluctuations where the asymmetry
was crucial for non-zero active force density. The wall breaks the symmetry
in the ``up'' and ``down'' directions and hence the hydrodynamic kernel
picks out a particular direction and does not depend on $(\wdn-\wup)$
when membrane fluctuations feel the wall.
The statistics for the noise $\tilfz$ in the presence of the wall 
follows from Eqns.\ (\ref{wallnoisecorr}).

The equal time correlation function can be calculated from
\bea
\pt \left(
\begin{array}{c}
            h(\kperp,\omega) \\  \psi(\kperp,\omega) \end{array} \right) =
-\left( \begin{array}{cc}
        \cG(\sigma k_{\perp}^2+\kappa
k_{\perp}^4)& \cG \kappa \hb k_{\perp}^2- \cGaw f  \\
                                        \Lambda \kappa \hb k_{\perp}^4 &
\Lambda \chi _0^{-1}k_{\perp}^2 
                                            \end{array} \right)
                                            \left(
                                        \begin{array}{c}
                                                h(\kperp,\omega)  \\
                                                \psi(\kperp,\omega)
                                                \end{array} \right) 
+ \left( \begin{array}{c}
                \tilfz(\kperp,\omega)\\ 
                i \kperp.\bf {f_{\psi}} 
                \end{array} \right). 
\eea
The correlation function is 
\be 
\label {actwallcorr} 
\langle h^{*}(\kperp)h(\kperp)\rangle = k_BT
\cG \left(\frac{(\tau_{\psi}^{-1}+{\tau_h^{-1}}^{w} -{\tau_{a}^{-1}}^{w} +
\frac{f^2 {\cGaw}^2}{\chi
_0^{-1}\cG})}{(\tau_{\psi}^{-1}+{\tau_h^{-1}}^
{w})({\tau_h^{-1}}^{w,e} + {\tau_a^{-1}}^{w})}\right), 
\ee
where
\be
{\tau_h^{-1}}^{w} = \cG(\sigma k_{\perp}^2 + \kappa k_{\perp}^4), 
{\tau_a^{-1}}^{w} = \cGaw \frac{\kappa \hb}{\chi_0^{-1}} f 
k_{\perp}^2. \nn 
\ee
${\tau_h^{-1}}^{w,e}$ has the same form as ${\tau_h^{-1}}^{w}$ with
$\kappa$ replaced by $\kappa_{e}$. 
Inserting physically reasonable values for the
parameters which enter these definitions (see below),
we verify that the right hand-side of the correlation
function is always positive. Thus, this correlation function
is calculated in a regime in which the active membrane
is in a {\em stable} steady-state phase.

At large length scales and in the
presence of a confining wall ($1/L_c  \ll  k_{\perp}  \ll 1/d$),
\be
\label{notenwallcorr}
\langle h^2(L)\rangle = \frac{k_B T}{\kappa \hb} 
\left[\frac{\epsilon^{\prime} 
{\chi_0}^{-1}}{\epsilon f}
+ \frac{f \epsilon}{D}\right] ln(L/a) \sim \frac{k_B T}{\Sigma} ln(L/a), 
\ee
for {\it tensionless} membranes and 
\be 
\label{tenwallcorr} 
\langle h^2(L)\rangle = k_B T
\left[\frac{\epsilon^{\prime}+ (f^2 \epsilon^2/D {\chi_0}^{-1})}
{\epsilon^{\prime} \sigma + (\epsilon f \kappa
\hb/{\chi_0}^{-1})}\right] ln(L/a) \sim \frac{k_B T}{\sigma^e}
ln(L/a), 
\ee
for tense membranes.  We have defined the quantities
$\epsilon = d {\wdn}^2 /(4 \eta)$ and
$\epsilon^{\prime} = d^3/(12 \eta)$ for notational
convenience.  Thus, the presence of the wall makes
tensionless active membranes less rough than they are
in bulk since the height correlation function scales
now logarithmically with $L$ instead of quadratically.
Note that this effect is {\em opposite} to that
predicted via the scaling arguments of Prost et al. in
Ref.\cite{prostmannbruinsma}.

The effective surface tension is then
\bea
\Sigma = {\kappa \hb} \Big{/} \left[\frac{\epsilon^{\prime} 
{\chi_0}^{-1}}{\epsilon f}
+ \frac{f \epsilon}{D}\right]. 
\eea 
This is an explicitly non-equilibrium effect, since $\Sigma
= 0 $ when $f = 0$. The induced surface tension also crucially depends
on the curvature-concentration coupling $\hb$ since it vanishes on
setting this coupling to zero. Therefore, both
non-equilibrium forces and curvature-concentration coupling are
essential for the stiffening of a tensionless, active membrane near a
wall. 

To calculate the collision length $L_c$ we set $h^2(L_c) \approx d^2$
in Eqns.\ (\ref{notenwallcorr}) and (\ref{tenwallcorr}). Thus,
\bea
 L_c(d) = \left\{ \begin{array}{ll}
                        a~exp\left(\Sigma d^2/k_B T \right) & \mbox{if
$\sigma =0$},
\\ 
                        a~exp \left(\sigma^e d^2/k_B T \right) & \mbox{if
$\sigma \ne 0$}.
                       \end{array}
                   \right.
\eea

The eigen-modes have frequencies
\bea
\omega_1,\omega_2 &=& \frac{1}{2} \left(D k_{\perp}^2 + \cG
\sigma k_{\perp}^2 +\cG \kappa {k_{\perp}}^{4} \right) \nonumber \\ 
&\pm&
\frac{1}{2} \left(\sqrt{({D k_{\perp}^2 + \cG \sigma k_{\perp}^2 +\cG
\kappa k_{\perp}^4) }^2 - 4(f \cGaw \hb \kappa \Lambda k_{\perp}^4 +
\cG D \kappa_{e} k_{\perp}^6 + \cG D \sigma k_{\perp}^4)} \right).\nn 
\eea
In the absence of curvature-concentration coupling, non-equilibrium
forces have no effect on the modes and the modes reduce to those of an
equilibrium membrane in the presence of a wall, within the linearized
theory.  For tensionless membranes with non-zero
curvature-concentration coupling, non-equilibrium effects become
apparent in the mode structure.  The modes to lowest order are
\bea 
\omega_1 &\approx&  \frac{f \cGaw \kappa
\hb}{{\chi_0}^{-1}} k_{\perp}^2 + \cG \kappa_{e} k_{\perp}^4 ,\nn \\ 
\omega_2 &\approx& D k_{\perp}^2 +
\frac{\kappa \hb (\kappa \hb \cG k_{\perp}^2 - f
\cGaw)}{{\chi_0}^{-1}} k_{\perp}^2.  
\eea
When membrane fluctuations are influenced by the wall, $\cG = d^3
k_{\perp}^2 / 12 \eta$ and $\cGaw = d {\wdn}^2 k_{\perp}^2/4 \eta$, to
lowest order. Far away from the wall
the modes reduce to those of a
non-equilibrium membrane in the bulk.

It may seem that for initially tensionless membranes
near a wall, the induced surface tension $\Sigma$ can
have either sign depending on the sign of the
curvature concentration coupling $\hb$. We will show
that for stability, it is imperative that $\hb$ is
positive for active membranes close to walls; a
negative $\hb$ leads to a negative $\omega_1$, and
thus an instability.  The requirement of positivity of
$\hb$ in the stable regime ensures that our
approximations of considering fluctuations at linear
order about an initial flat steady-state phase are
self-consistent.

We now present a physical argument for why pumping
activity should lead to the {\em stiffening} of
active, tensionless membranes near repulsive walls.
The calculations of Section IV indicate that the
eigen-values governing the relaxation of a tense,
equilibrium membrane near a wall are $\sigma d^3
k_{\perp}^4/(12 \eta)$ and $D k_{\perp}^2$. For an
active, tensionless membrane near a wall these
eigenvalues are $\frac{\kappa \hb f d {\wdn}^2
k_{\perp}^4}{4 \eta \chi_0^{-1}}$ and $D
k_{\perp}^2$.  The coefficient of the slower mode for
the active case ($O(k_{\perp}^4)$) can be written as
$\frac{\kappa \hb \epsilon f}{\epsilon^{\prime}
\chi_0^{-1}} d^3/(12 \eta)$ and compared to the
expression for the slowest relaxation mode for tense,
equilibrium membranes. This comparison suggests that
an apparent surface tension ($\frac{\kappa \hb
\epsilon f }{\epsilon^{\prime} \chi_0^{-1}}$) is
induced due to purely non-equilibrium effects, even
for a membrane with vanishing bare surface tension.

In the linearized theory, the induced surface tension
obtained using the above argument and the surface
tension $\Sigma$ obtained from the coefficient of the
logarithmic term in the calculation of the height
roughness coincide.  It is thus apparent that the
presence of the wall smoothens a tensionless, active
membrane by making it relax faster than its
tensionless equilibrium counterpart (whose slowest
mode is $O(k_{\perp}^6)$).

Why does an active membrane relax faster near the
wall? Height field fluctuations are predominantly
governed by the $\psi$ field (the second term in
Eq.\ \ref{actwallheight}) because this term is
$O(k_{\perp}^2)$ in contrast to the other terms which
are of higher order in wavenumber. Thus, this equation
to lowest order in wavenumber is $\pt h = \cGaw f \psi$.
Since the $\psi$ field relaxation is fast
compared to the height field, we may assume that, over
the larger time-scales of interest to us in studying
the dynamics of the height field, $\psi$ relaxes
instantaneously to $\psi_s =-\frac{ \kappa
\hb}{{\chi_0}^{-1}} k_{\perp}^2 h$.
The effective height field equation now becomes
\be
\pt h = - \frac{f \epsilon \kappa \hb}{{\chi_o}^{-1}} k_{\perp}^4 h.
\ee 
Note that the coefficient of the $k_{\perp}^4$ term in
the above equation can be rewritten to obtain the
induced surface tension $\Sigma$, as explained in the
previous paragraph.  

The above argument invokes and crucially requires the
presence of the wall, activity and
curvature-concentration coupling. In physical terms,
height field fluctuations are accompanied by fast
relaxation of the protein density field $\psi$. This
relaxation is to a steady state value determined by
the pump diffusion coefficient and
curvature-concentration coupling. The pumping activity
is then inhomogeneously distributed along the membrane
surface. To see the physical consequences of this
inhomogeneous distribution, consider a fluctuation
creating a region of positive curvature.  If $\hb$ is
positive, the number of ``up'' pumps in the region of
positive curvature tends to decrease with time and the
decrease in the pumping activity causes a smoothening
of the membrane.  However, if $\hb$ is negative, the
region of positive curvature attracts more ``up''
pumps which further bootstraps the height fluctuation,
leading to an instability.

The explicit expressions obtained above also permit us
to estimate the collision length for tensionless
active membranes in the presence of repulsive walls.
Using estimates given in\cite{sriramprostmannbass} for
various physical quantities ($\kappa = 10^{-20}~
Joules, \hb = 5~nm, f = 10^{-12}~N,d = 1~micron,
{\chi_0}^{-1} = 10^{-21}~Joule~m^2$), we find that the
induced surface tension $\Sigma \sim
10^{-7}-10^{-8}~N/m$. This translates to collision
lengths $L_c$ of the order of $10^5 -10^6$ times
molecular scales. Assuming molecular length scales of
order a nanometer, this calculation would then predict
collision lengths of the order of millimeters, for 
membranes with physical parameters in the above 
range.

\section{Conclusions}\label{conclu}

This paper has presented a calculation of the
linearized hydrodynamics of two-component membranes
including the effects of curvature-concentration
coupling, specifically in the context of a simple
model for a biological membrane with protein and lipid
constituents. We have studied in some detail the case
in which such membranes are placed close to as well as
far away from a confining repulsive wall.  We have
considered the case in which such proteins are
``active'', in which case they exert non-equilibrium
forces on the solvent, as well as the case of
``inactive'' pumps, corresponding to thermal
equilibrium.

The calculations we report are for a model which
shares many similarities with that of
Ref.\cite{sriramprostmannbass}.  Manneville {\it et
al.} consider the more general case of a membrane with
a finite permeability whereas we work with an
impermeable membrane throughout.  In addition, we
neglect the possible effects of curvature on
activity.  However, the relatively simpler structure
of our model enables us to establish a connection with
equilibrium results in a more transparent fashion.  We
work in a framework in which boundary conditions are
explicitly implemented.  This is especially of use
when we consider the linearized theory of fluctuations
of an active membrane near a wall. The results we
report for the linearized hydrodynamics of active and
inactive two-component membranes near walls are new,
as are our results for the equilibrium case far away
from confining walls.

We find, far away from the walls and consistent with
the results of Manneville {\it et al.}
\cite{sriramprostmannbass}, that the presence of
active pumps on the membrane does not change its
roughness. In the case of tensionless membranes,
fluctuations can be described in terms of an effective
temperature which is shifted from the thermodynamic
temperature by non-equilibrium effects due to pumping.
These terms depend on the quantity $(\wdn-\wup)$, a
measure of the degree of asymmetry in the positions of
the force-centres. We reiterate the observation of
Manneville {\it et al.} in Ref.
\cite{sriramprostmannbass} that this asymmetry is
essential for non-equilibrium effects to be visible in
the membrane dynamics.

We find that the activity of the pumps within the
{\em stable} regime  appears to {\em
smoothen} out height fluctuations in the presence of
the wall since height correlations scale only
logarithmically with $L$ for both tense and
tensionless membranes. This smoothening out within the
linear theory presumably competes with the possible
enhancement of fluctuations, if any, which might arise
in a theory which incorporates the effects of
non-linearities. We have provided a physical
interpretation of this effect in this paper and
obtained estimates for the appropriate collision
length. We find that this collision length can be
significantly larger than typical length scales for
biological vesicles, indicating the relevance of the
physical effects we study here to a complete
description of the hydrodynamics of active
pump-membrane systems. It is important to note that
this smoothening relies crucially on the existence of
curvature-concentration coupling terms, neglected in
previous work on the near-wall case. These results
suggest that earlier ones based on scaling arguments
which predicted an {\em enhancement} in the roughness
may need to be reevaluated in the light of this work.

We have also calculated the ``crossover'' function
$\cGaw$ and linearized mode structure in the presence
of the wall.  While the positional asymmetry in the
placement of force-centres was crucial for active
membrane fluctuations in bulk, we point out that
non-equilibrium effects are apparent even for
symmetrically placed pumps in the presence of a wall.
This unusual, although not totally unexpected, result
does not appear to have been noticed before.

We also find that the requirement of stability for
active membranes near a wall constrains $\hb$, the
coefficient coupling concentration with curvature, to
be positive. This can have interesting consequences.
Consider two active membranes, one with positive $\hb$
and the other with negative $\hb$, fluctuating in the
bulk. Provided $\hb {\cal P}$ for each lies in the
window of stability in the bulk, these membranes are
stable.  Now bring both close to a wall. The membrane
with positive $\hb$ will remain stable. The
fluctuations of the membrane with negative $\hb$
however, will turn unstable; its final fate will be
decided by the non-linear terms we have omitted from
our discussion.

Because our calculation is an explicitly {\em linear}
one, we cannot address issues such as the relevance of
such non-linear terms.  The arguments of Prost,
Manneville and Bruinsma which predict an anomalous
roughening of active membranes near walls rely on
identifying the role of such non-linear terms and
estimations of their importance via scaling
arguments.  It would be interesting to construct
similar arguments for the model studied here,
particularly since our results in the linear case
indicate that fluctuations of an active membrane near
a repulsive wall can be {\em smoothened} in some
regimes, rather than roughened relative to the
equilibrium case. Further work along these lines is
currently in progress.

\section*{Acknowledgments} 

We are grateful to Sriram Ramaswamy for a critical
reading of the manuscript as well as several
enlightening discussions on these and related
matters.  We also thank Y. Hatwalne for urging us to
clarify the physical mechanisms underlying our
results.

\newpage

\section{Appendix}

\subsection{ A note on the impermeability of the membrane} 

In this sub-section we present an argument, following
Manneville {\it et. al.\/}, to justify the assumption
of membrane impermeability.  Had we incorporated
permeation due to activity in our calculations,
active-permeative and active-hydrodynamic terms would
appear as $(\lambda_p - \frac{k_{\perp}
({\wup}^{2}-{\wdn}^{2})}{8 \eta})f \psi$, ($\lambda_p$
is the permeation coefficient)
\cite{sriramprostmannbass}.  The cross-over length
scale from the hydrodynamic to the permeative regime
occurs when $k_{\perp} = \frac{8 \eta \lambda_p}{w
(\wup-\wdn)}$, {\it{i.e.}}  $l_{\perp} = \frac{\pi w
(\wup-\wdn)}{4 \eta \lambda_p}$ ($w$ is the size of
the pump).  With $\lambda_p = 10^{-12} m^3/N.s $,$\eta
= 10^{-3} kg/m.s$ and $w = 5.10^{-9} m$
\cite{sriramprostmannbass}, this crossover length
scale is estimated to be of the order of a
centimeter.  The length scales of concern to us are in
the micron and sub-micron range
\cite{sriramprostmannbass}. Thus, at these length
scales, effects due to permeation can be ignored. These
arguments can be extended to estimating the role
of passive permeative {\it vs.\/} hydrodynamic modes of
relaxation; we conclude that at the length scales of
interest to us, the assumption of membrane impermeability
remains valid.

In the presence of a wall, active permeative terms
would dominate if $\lambda_p > \cGaw$. This means that
the length scale above which permeation dominates is
$\sqrt{\frac{\pi d {\wdn}^{2}}{2 \eta \lambda_p}}$.
This length is approximately $40$ microns when the
membrane-wall distance $d=1$ micron. Hence, in the
regimes of interest to us, permeation can be ignored
and relaxation is purely hydrodynamic. Hence, we
assume $\lambda_p = 0$ to derive our results.   

\subsection{A modified model for active membranes}
We present here results for a model in which proteins are located
asymmetrically across the bilayer with their centers of mass displaced
above or below the bilayer midpoint (see Fig. 3). The head and tail of
the pumps are then located at distances $w_l$ and $w_s$ respectively
from the midpoint of the bilayer. Note that in this model, the ``up''
and ``down'' pumps are related to each other by reflection along the
$z=0$ line unlike the model of Fig. 2. This model can be used to
describe situations in which proteins are attached to planar bilayer
membranes which are symmetric on both sides, unlike vesicles in which
the inner leaf and outer leaf need not be symmetrically preferred by
the protein.

When pumps are active on such a membrane in bulk, the force
density due to the pumps acting on the fluid has the form \bea
f_{pump-fluid} = f[\delta(z-w_l)\nup-
\delta(z+w_s)\nup-\delta(z+w_l)\ndn+\delta(z-w_s)\ndn], \eea where
$\nup$ and $\ndn$ are the local number of ``up'' and ``down'' pumps
respectively.  On repeating the calculations done in Section V, we
find that the $z$-component of the hydrodynamic velocity has the same
form as shown in Eq.\ (5.7) with $\Omega(k_{\perp}, \wup,\wdn)$
replaced by $\Omega^{\prime}(k_{\perp},w_l,w_s)= -e^{k_{\perp}
w_s}(1+k_{\perp} w_s)+e^{k_{\perp} w_l}(1+k_{\perp} w_l)$, which
vanishes when $w_l=w_s$ ({\it i.e.} for symmetrically placed
proteins). Correlation functions and relaxation times can be read off
from the corresponding expressions in Section V by identifying $\wdn$
and $\wup$ with $w_s$ and $w_l$ respectively.  For example,
\be
\langle h^{*}(\kperp,t)h(\kperp,t)\rangle =
\frac{k_BT}{\eqker} \left(
\frac{(\tau_{\psi}^{-1}+\tau_{h}^{-1}-\tau_{a}^{-1} +\frac{f^2
\Omega^2}{\eqker \chi_0^{-1}})}{(\tau_{\psi}^{-1}
+\tau_{h}^{-1})({\tau_{h}^{-1}}^e+\tau_{a}^{-1})} \right), \nn
\ee
where the quantities $\tau_{\psi}^{-1},\tau_{h}^{-1}$ have been
defined earlier. ${\tau_{h}^{-1}}^e$ has the same form as
$\tau_{h}^{-1}$ with $\kappa$ replaced by $\kappa_{e}$. The relaxation
time
\be
\tau_{a}^{-1} =
\frac{1}{\eqker}\frac{\kappa \hb}{{\chi_0}^{-1}} f
\Omega(k_{\perp},w_l,w_s) k_{\perp}^2. \nn
\ee
We note here that this identification is purely notational since
$(\wdn,\wup)$ and $(w_s,w_l)$ physically represent different lengths
in the two models.

When the calculations are repeated for such a model in the presence of
a repulsive wall, we find that the hydrodynamic velocity acquires an
additional contribution from the $\phi$ (total protein density)
field. In this linearized calculation we assume that the
compressibility associated with the $\phi$ field is very large and
ignore its dynamics.  However, we do not expect this contribution to
alter the long wavelength fluctuations of the active membrane near the
wall significantly.

\newpage

\begin{center}
{\bf Figures}
\end{center}


\begin{figure}
\myfigure{\epsfysize1.7in\epsfbox{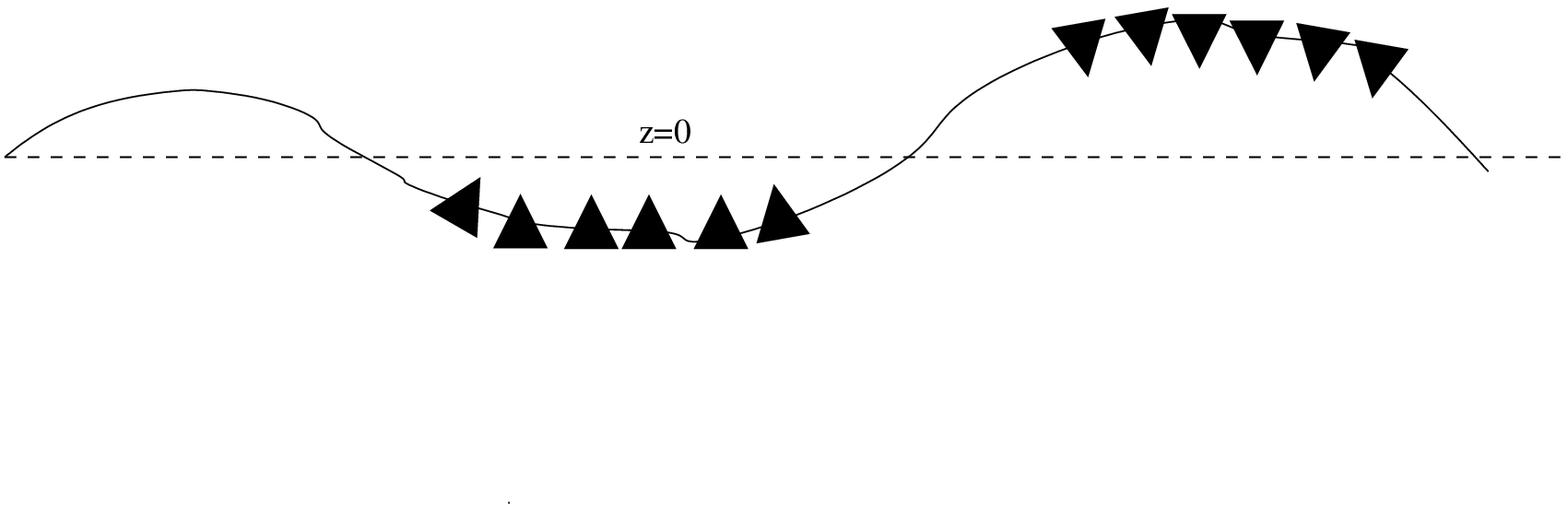}}{Fig. 1. A schematic diagram
of the two kinds of proteins, represented as ``up'' and ``down''
triangles. These affect the local curvature of the membrane by bending
it towards or away from the local normal. The bilayer mid-plane lies
at $z=0$.}
\end{figure}

\begin{figure}
\myfigure{\epsfysize1.7in\epsfbox{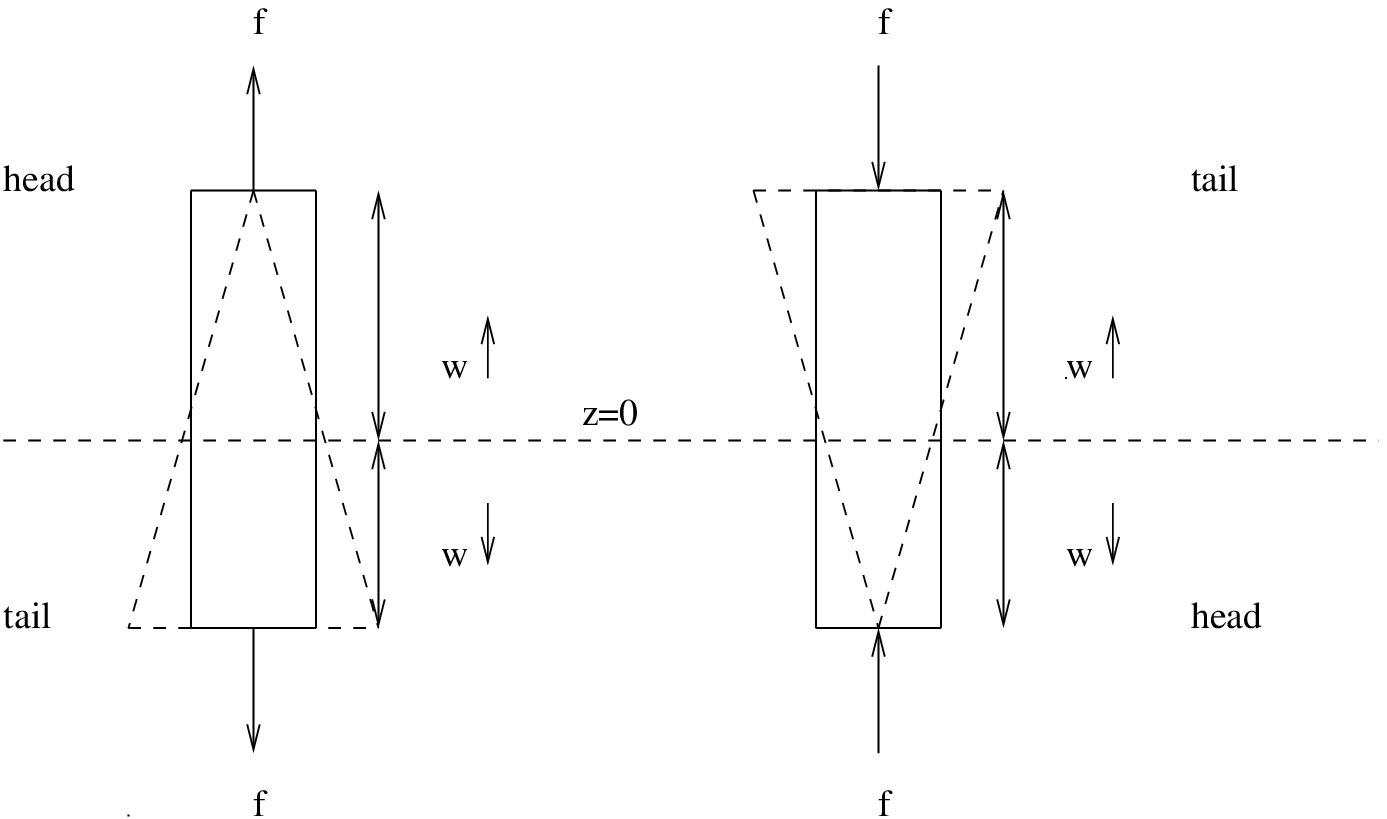}}{Fig. 2. The asymmetric
dipole model used to describe the activity of the proteins. The
force-centres are places at distances $\wup$ and $\wdn$ from the
bilayer midpoint. The centers of mass of the pumps are displaced above
the bilayer midpoint.  The superposed ``up'' and ``down'' triangles
indicate the underlying shape asymmetry of the two different kinds of
pumps.}
\end{figure}

\begin{figure}
\myfigure{\epsfysize1.7in\epsfbox{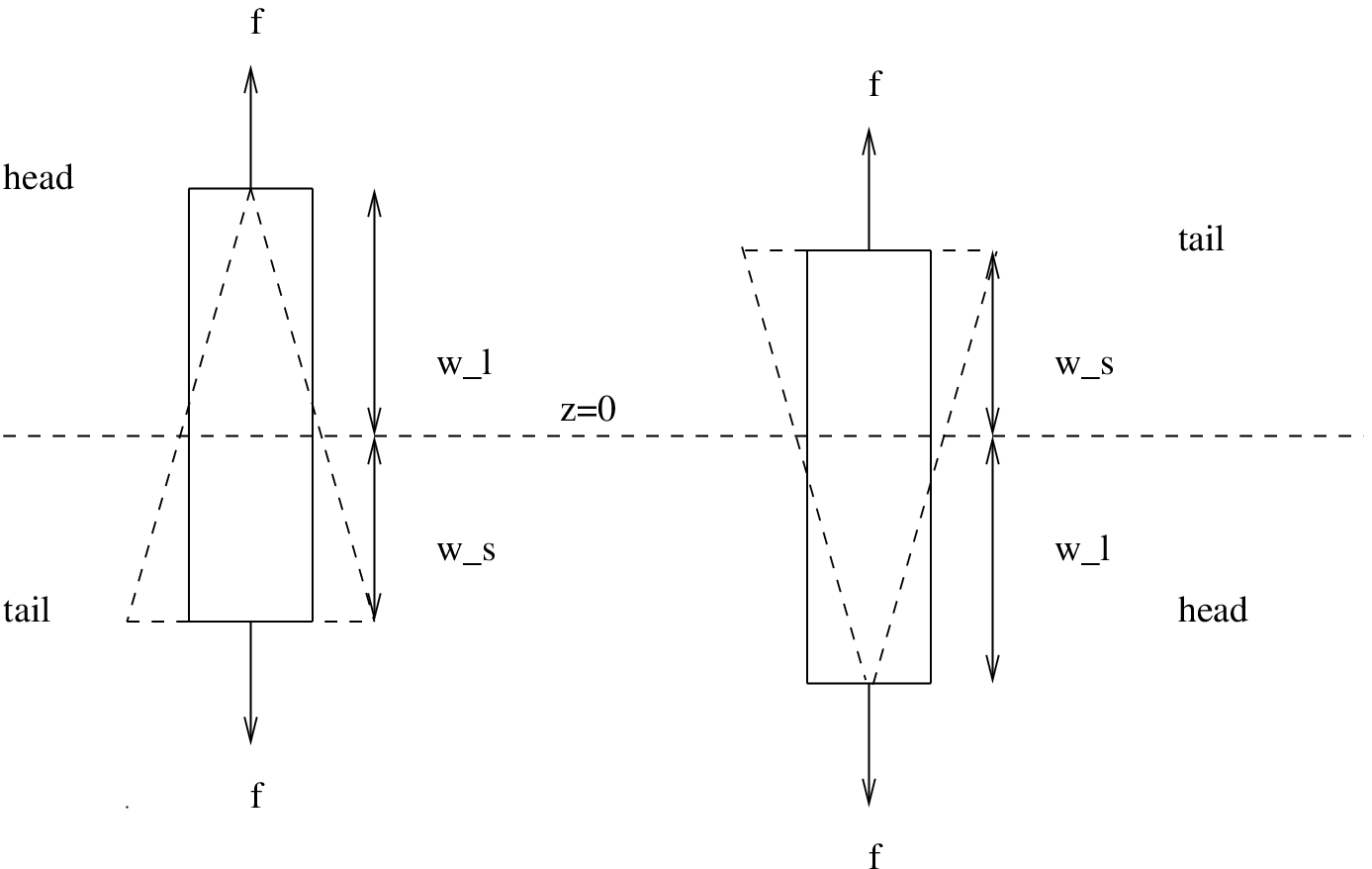}}{Fig. 3. The asymmetric
dipole model of the pumps described in the Appendix.  The centers of
mass of the pumps are displaced above or below the bilayer
midpoint. The distance of the furthest force-centre from $z=0$ is
$w_l$ and the distance of the nearest force-centre from $z=0$ is
$w_s$.}
\end{figure}


\begin{references}
\bibitem{compress} At large enough length scales, the membrane 
is compressible as shown in\cite{cailubenskypre}, due to area
stored in membrane fluctuations at short length scales. We have
assumed $\rho \simeq \rho_0$, implying that such density fluctuations
should be strongly suppressed at the length scales of relevance to
our calculation.
\bibitem{comment} The RHS of 
Eq.\ (\ref{actcorr}) must be positive. Since
other quantities are explicitly positive, we require that $\tau_h^{-1}
-\tau_a^{-1} > 0$, implying $\hb {\cal P} w / \chi_0^{-1} \ge -1$.
We see that the values of $\hb {\cal P}$ satisfying the stability
requirement automatically satisfy the condition of the positivity of the
correlation function.  We thus conclude that the correlation function
becoming negative can be associated with the onset of an instability in
the system.
\bibitem{noneqkernel} The complete expression for the non-equilibrium
kernel is 
{\small{
\bea
\cGaw(k_{\perp},d,\wup,\wdn) &=& 
\frac {1}{\eqker (-1+e^{2 k_{\perp} d}+ 2 k_{\perp} d - 2k_{\perp}^{2} d^2)} [-2 \cosh (2
k_{\perp}d - k_{\perp} \wdn)+k_{\perp} \wdn 2 \sinh (k_{\perp} \wdn) \nn \\
&+& 2 (1+2 {k_{\perp}}^2 d^2 -
2 {k_{\perp}}^2 d \wdn) \cosh (k_{\perp} \wdn) + 2 (2 k_{\perp} d - k_{\perp} \wdn) \sinh (k_{\perp}
\wdn) \nn \\ 
&+& 2 e^{-k_{\perp} \wup} (1 + k_{\perp} \wup) \cosh (2 d k_{\perp}) + 2 (1+ 2 {k_{\perp}}^2 d^2)
e^{- k_{\perp} \wup}] \nn
\eea}}
\end{references}
\end{document}